\documentstyle[multicol,aps,pre,epsf]{revtex}

\def\s{\sigma}
\def\t{\theta}
\def\tb{\bar\theta}
\def\D{\Delta}
\begin{document}

\title{Persistence in systems with algebraic interaction}

\author{Iaroslav Ispolatov}
\address {Department of Physics,
McGill University, 3600 rue University, Montr\a'eal,
Qu\a'ebec, H3A 2T8, Canada }
\address{
and}
\address{ 
Chemistry Department, Baker Laboratory, Cornell University, Ithaca,
NY 14853, USA}
\date{\today}

\maketitle
\begin{abstract}
Persistence in coarsening 1D spin systems with a power law 
interaction $r^{-1-\sigma}$ is considered.
Numerical studies indicate that for sufficiently large values of  
the interaction 
exponent $\sigma$ ($\sigma\geq 1/2$ in our simulations), 
persistence decays as an algebraic 
function of the
length scale $L$, $P(L)\sim L^{-\theta}$.
The Persistence exponent $\theta$ is found to be independent on
the force exponent $\sigma$ and close to its
value for the extremal ($\sigma \rightarrow \infty$) model,
$\bar\theta=0.17507588\ldots$.
For smaller values of the force exponent 
($\sigma< 1/2$), finite size effects prevent the system from 
reaching the asymptotic regime.
Scaling arguments suggest that in order to avoid significant 
boundary effects for small $\sigma$, the system size should
grow as ${[{\cal O}(1/\sigma)]}^{1/\sigma}$. 


\end{abstract}

\begin{multicols}{2}

Coarsening dynamics of one-dimensional systems with a  power-law
$V(r)\sim r^{-\s-1}$ interaction between spins
has recently been studied by Lee and Cardy \cite{lc}, and Rutenberg and Bray
\cite{rb}. It had been established that 
after quenching  from a high-temperature 
disordered phase to $T=0$ these systems 
develop a domain structure characterized by a 
single lengthscale $L(t)$. A naive 
argument based on the law of motion for domain walls, 
$\dot L\sim L^{-\s}$ (where $L^{-\s}$ 
is a typical force between domain walls), produces an asymptotically 
correct time dependence of $L$, 
\begin{equation}
\label{L}
L(t)\sim t^{1/1+\s}.
\end{equation}
Other 
properties of this system, including correlation functions and 
domain size distribution, have been studied in \cite{rb} as well.

In this paper we shall look at another facet of 
1D phase-ordering systems
with a power-law interaction: what fraction $P$ of spins have never 
changed sign up to the time $t$? Or, equivalently, what fraction of the 
space has never been crossed by a domain wall?
Such a property of coarsening
systems is usually called persistence and has recently 
become a 
major subject of research in statistical physics \cite{bdg,dhp,b1,b2,b3}.
Let us briefly review some known results in this field relevant to our 
problem.
In \cite{bdg} the exact solution was found for persistence in an ordering 
system
with extremal dynamics. The extremal dynamics limit is reached by 
formally setting
$\s\rightarrow \infty$, which means that interactions 
become infinitely short-range. 
In this limit, coarsening proceeds
by consecutive shrinking and disappearance of 
the current smallest 
domains in the system, while  other domain boundaries remain virtually 
motionless. 
It was established in \cite{bdg} that persistence at a stage of 
evolution 
when the average domain size is $L$ is proportional to $L^{\tb}$, 
where the exponent $\tb = 0.17507 \ldots$ is the solution of 
an implicit integral equation.

In \cite{dhp} persistence exponents have been calculated for  
coarsening 1D Potts models with Glauber dynamics. For the 2-state Potts (Ising)
model, persistence decays as $t^{-\t}$, $\t=3/8$, or in terms of the average 
domain size $L$, $P(L)\sim L^{-3/4}$.

The following conclusion can be drawn from a comparison of
persistence exponents for extremal and Glauber dynamics. 
Extremal dynamics is more efficient in preserving persistence, since 
the motion of domain walls is always directed towards their ultimate 
annihilation partners,
while in the case of Glauber dynamics, domain walls perform
random walks and sweep through a larger amount of space, which otherwise
could have remained persistent.
The extremal dynamics exponent $\tb$ sets a lower
bound on persistence exponents for systems with a finite force exponent $\s$.
It is easy to visualize a scenario when a domain wall first moves away from 
its ultimate annihilation partner, and then, after the stronger force source 
disappears, it turns back. Such events result in spin flips on
parts of the line that belong to a surviving domain and would have been 
left untouched in the extremal dynamics case.

The results presented below suggest that this lower boundary
$\tb=0.17507 \ldots$ is in fact the exact value of the persistent exponent 
for arbitrary $\s>0$.

Let us formally introduce our model:
We consider coarsening of the 1D 2-state spin system with a long-range
ferromagnetic Hamiltonian: 
\begin{equation}
H = {-4\over\s}\sum_{i > j}{s_is_j \over {(x_i-x_j)}^{\s+1}}.
\end {equation}
After quenching  from a high-temperature random phase to $T=0$,
coarsening
dynamics for this system is determined by the motion of domain walls,
governed by the Langevin equation.
The velocity of a 
wall 
is equal to the sum of pairwise forces from other walls, with walls of the 
same signs repelling and walls of the opposite signs attracting each other:
\begin {equation}
\label{f}
{d r_i \over dt} = \sum_{j \neq i} {(-1)}^{i+j}
{\mathrm sign}(r_i-r_j) F_{ij}
\end {equation}
\begin {equation}
F_{ij} = {1 \over {|r_i-r_j|}^{\s}}.
\end {equation}
When the adjacent walls meet, they annihilate.
As we mentioned, the degree of coarsening is uniquely characterized by
a typical domain size $L(t)\sim t^{1/(1+\s)}$. We measure the fraction 
of space $P(L)$
that has never been crossed by a single domain wall as a function of this
lengthscale $L(t)$. 
We perform molecular dynamics simulations of the model for the 
following values of the force exponent:
$\s=3/2,\: 5/4,\: 1,\: 3/4,\: 1/2,\: 1/4$.
Each run starts with a system consisting of 
$N_0= 100000$ domain walls; results for each $\s$ are averaged over 20
initial configurations.
Open boundary conditions with no replicas added to 
the boundaries are used. 
To speed up the evaluation of forces, a 1D multipole expansion
has been performed, and terms of up to quadrupole order
were taken into account \cite{ik}.
\begin{figure}
\centerline{\epsfxsize=8cm \epsfbox{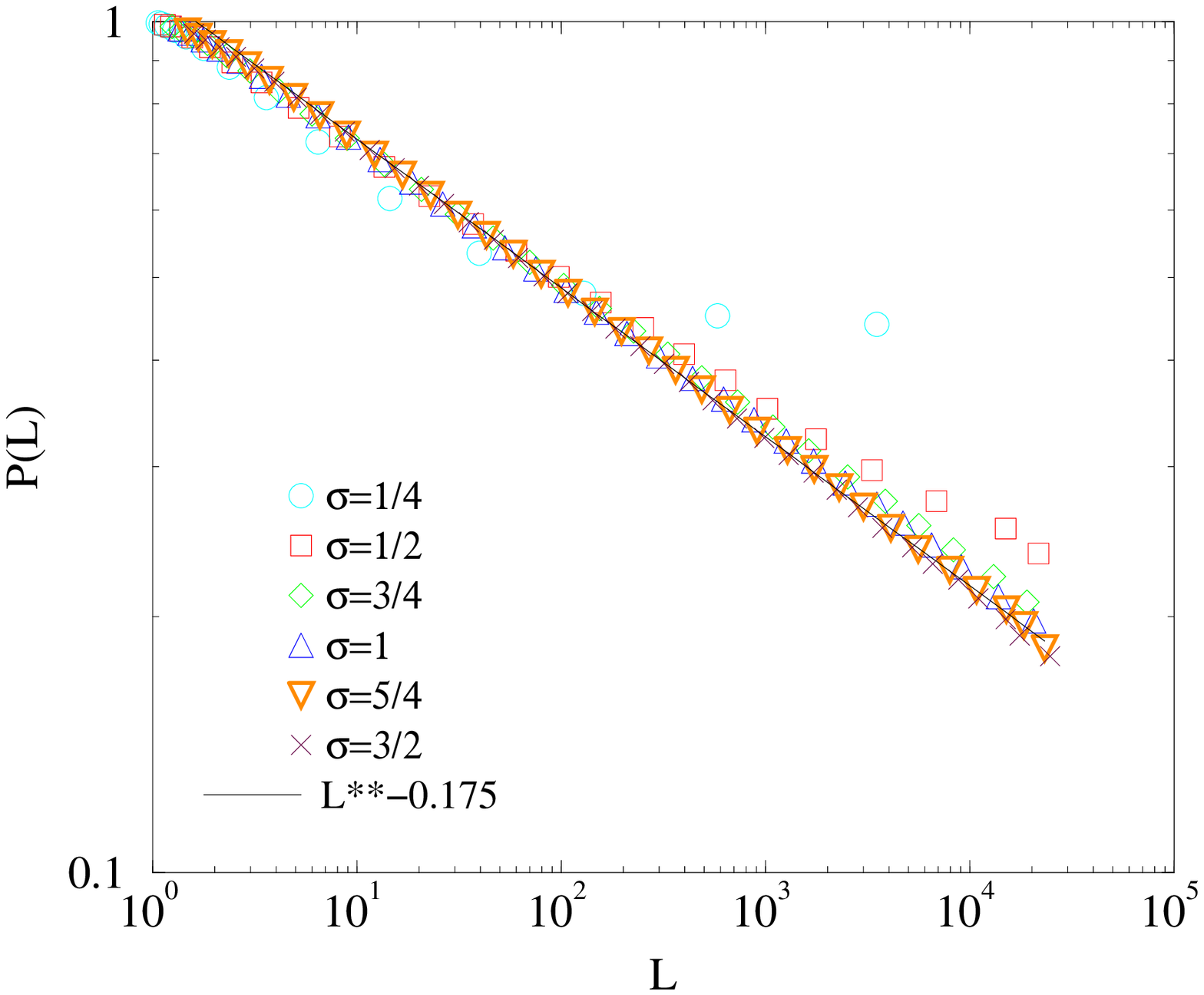}}
\noindent
{\small {\bf Fig.~1}. 
The log-log plot of persistence $P(L)$ vs. average domain size $L$
for various force exponents $\s$. The straight line corresponds to
$P(L)\sim L^{-\tb}$.}
\end{figure}
The results for persistence as a function of the average domain length $L$ are 
presented in log-log form in Fig.~ 1. Except for small force exponents 
($\s=1/4$ and later evolution stages for $\s=1/2$) all the curves collapse 
at
a line with a slope $\approx -0.175$, which corresponds to $\s=\infty$ 
extremal model.

Statistical error bars are shown in  Fig.~2 for a single set of data
($\s=5/4$).

Our simulations suggests that scaling of persistence, corresponding 
to $\s=\infty$, is 
valid for all other not very small $\s$.  The following asymptotic 
argument helps to understand
why this is so. At any current moment of time, persistent spins are
mostly contained in the domains that were expanding at almost all 
previous stages of coarsening;
{\it i.e. } these domains were larger than the average at those stages.
If one of such large domains is surrounded by two small neighbors, it would
most probably grow outwards, and no spin flips, additional to those inevitably 
caused
by directed coarsening itself, would happen. The situation may be 
different if 
two or three big domains are adjacent to each other: their 
domain walls may wonder and get inside the territory of the future 
survivor, causing some 
excessive spin flips.
\begin{figure}
\centerline{\epsfxsize=8cm \epsfbox{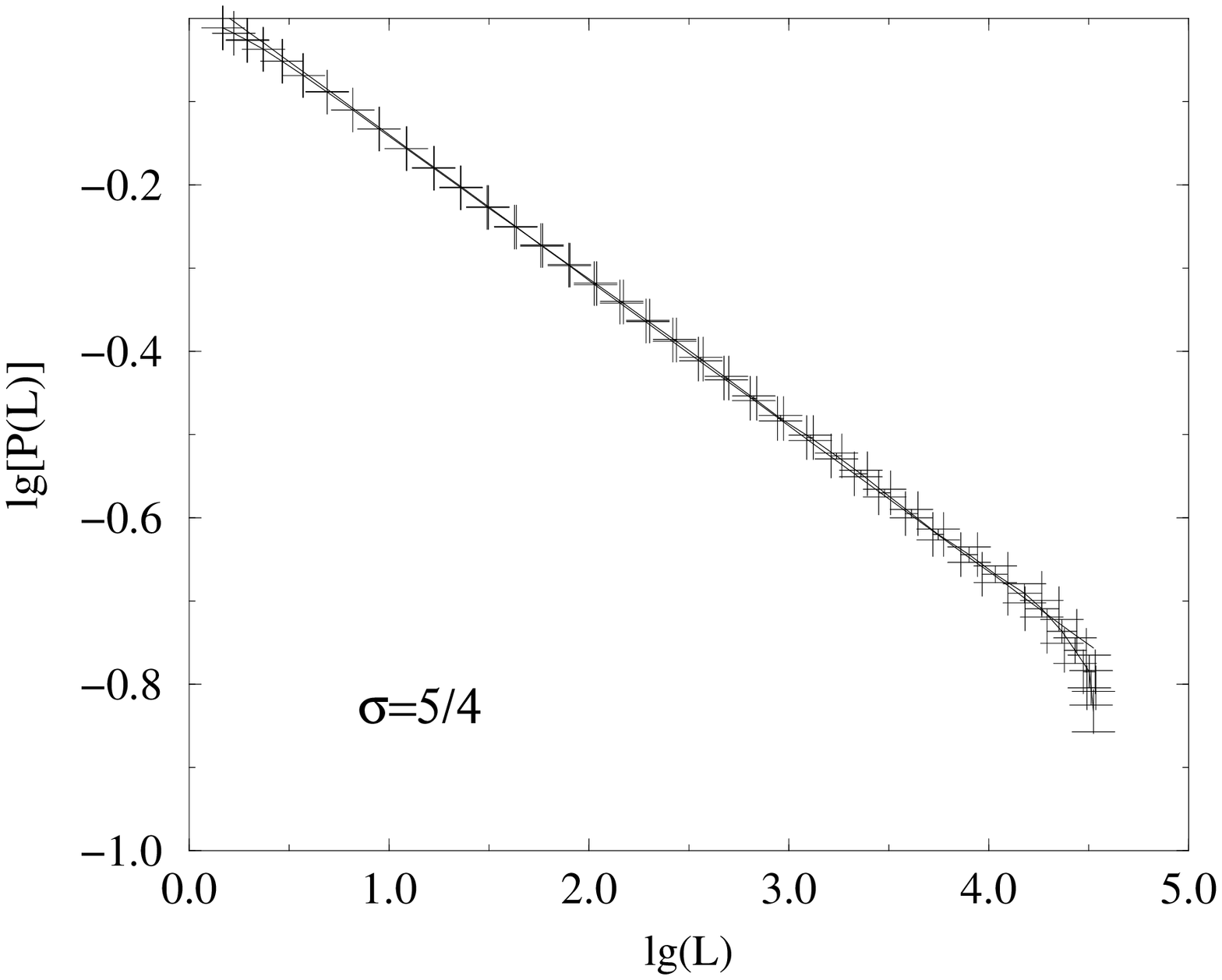}}
\noindent
{\small {\bf Fig.~2}. 
The log-log plot of persistence $P(L)$ vs. average domain size $L$
force exponents $\s=5/4$ with statistical error bars. 
The straight line corresponds to $P(L)\sim L^{-\tb}$.}
\end{figure}
We can estimate the characteristic scale of  such a persistence-loosing event.
A typical distance $\D L$ that a wall of domain of size $L_0$, surrounded by
a group of domains of similar sizes,  travels during time $t$ is 
\begin{equation}
\label{dl}
\D L\sim L_0 - (L_0^{(1+\s)}-t)^{1/1+\s}\approx L(t) 
[{L(t)\over L_0}]^{\s}.
\end {equation}
Here $L(t)\sim t^{1/1+\s}$ is the average domain size at time $t$.
For positive $\s$, $\D L$ becomes negligible
compared to $L$ when $ {L(t)\over L_0} \gg 1$, hence the number of spin flips
additional to those present in extremal dynamics coarsening scenario 
becomes negligible.
Another conclusion that follows from Eq.~(\ref{dl}) is that for small $\s$,
the crossover time to $P(L) \sim L^{\tb}$ scaling must be 
larger since the 
system must develop structure that includes sufficiently 
large domains.

However, besides long intitial transitional times, 
there is another reason for the breakdown of scaling for small
$\s$ that we observed in our simulations. Let us first consider the 
opposite to $\s=\infty$ 
case of $\s=0$. In this limit forces, are distance-independent,
and domain wall dynamics (\ref{f}) is described by the equation
\begin {equation}
\label{fn}
{d r_i \over dt} = \sum_{j \neq i} {(-1)}^{i+j}
{\mathrm sign}(r_i-r_j).
\end {equation}
Since domain walls come in pairs, the sum in Eq.~(\ref{fn}) is equal to 
$\pm1$. That means that all walls have the same constant velocity with 
odd-number walls moving to the left and even-number moving to the right. 
The whole system becomes a collection of independently collapsing and 
growing domains. This clearly violates the scaling (\ref{L});
in fact the $\s=0$ system has two 
lengthscales, $L_0-2vt$ and $L_0+2vt$,
where $L_0$ is the average initial domain length 
and $v=1$ is the velocity of
domain walls.
For an exponential distribution of initial domain sizes,
$W(L_0)=\exp(-L_0)$, persistence can be expressed as 
\begin{equation}
\label{0}
\tilde P(t)={\int_{0}^\infty W(x) dx+ \int_{2vt}^\infty W(x) dx
\over 2 \int_{0}^\infty W(x) dx}={1+\exp[-2t] \over 2}
\end{equation} 
Systems with few particles and  small $\s$  coarsen almost according to
the $\s=0$ scenario: particles across the whole system feel the presence
of the boundary. Odd and even -number walls tend to move predominantly 
to the left and right,
respectively, independent of the position of their nearest neighbors.

To probe whether the deviation from $P(L) \sim L^{\tb}$ 
scaling in persistence 
behavior is caused by $\s=0$ finite size effects, we do the following 
measurements. First, for the system of the same initial size 
$(N=10^5)$ we plot the average domain length $L(t)$ as a function 
of time and compare it 
to the $L\sim t^{1/1+\s}$ prediction.
\begin{figure}
\centerline{\epsfxsize=8cm \epsfbox{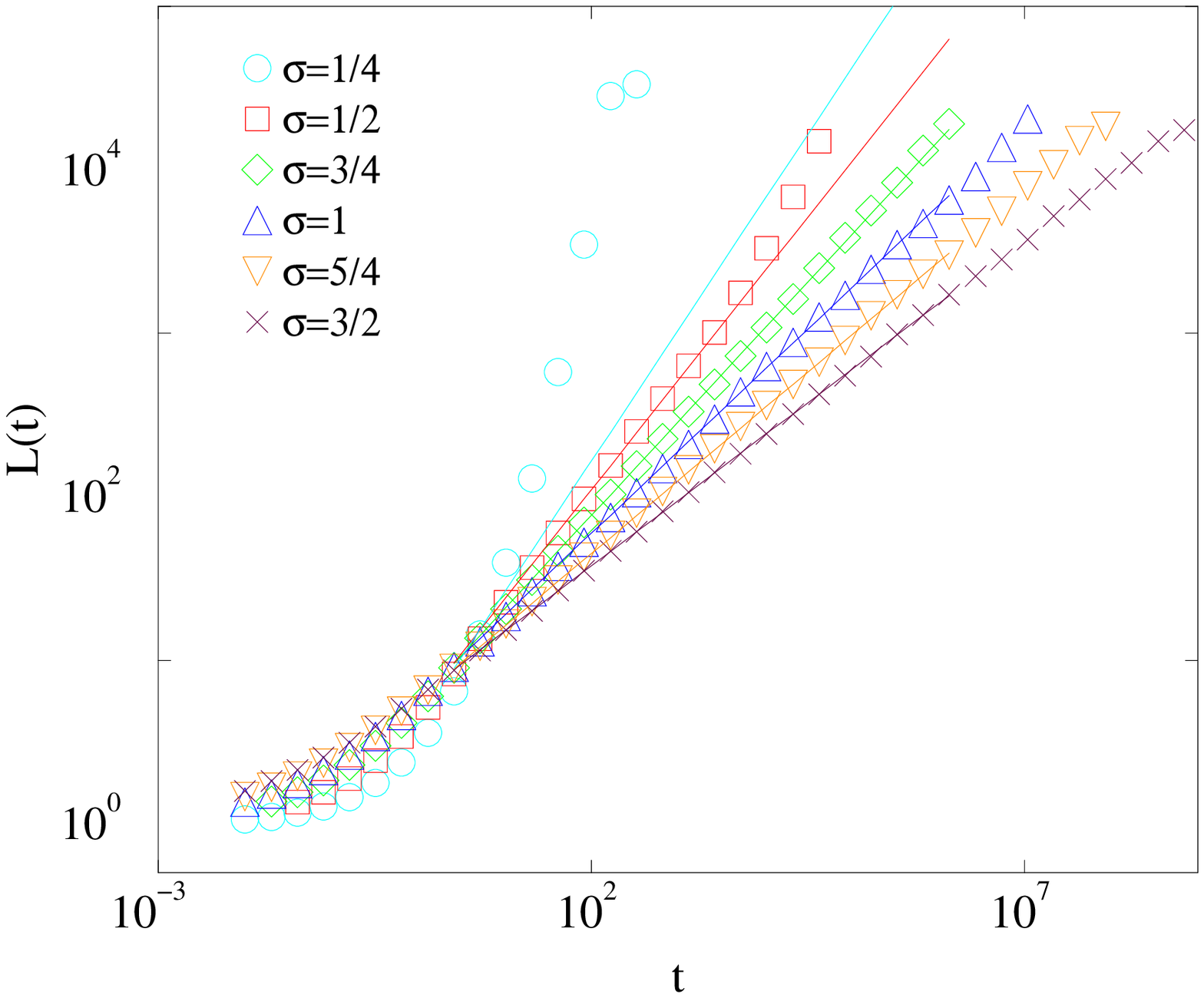}}
\noindent
{\small {\bf Fig.~3}. 
The log-log plot of average domain size $L(t)$ vs. time
for various force exponents $\s$. 
Straight lines correspond to scaling predictions, 
$L(t)\sim t^{1/1+\s}$.}
\end{figure}
Results for this simulation are presented in Fig.~3.
One can see that the system with $\s=1/4$ is never in 
scaling regime (\ref{L}),
system with $\s=1/2$ behaves according to  (\ref{L}) only up to some 
intermediate stage of evolution. For all other force exponents $\s>1/2$, 
for a certain period of evolution after short 
transitional time,  typical domain sizes scale according to (\ref{L}). 
Another check of whether a system feels the presence of the boundaries and
therefore crosses over to the $\s=0$ coarsening regime, 
is to measure directly the 
fraction of domain walls $B(L)$ that move opposite 
to the direction prescribed
by boundary effects. 
In Fig.~4 we plot the fraction of even-number domain 
walls moving to the right and odd-number domain walls moving to the left
for the systems initially consisting of the same number of domains,
$N=10^5$.
For finite $\s>0$ and a truly infinite system this fraction should be 
equal to $1/2$, for $\s=0$ it should be 0.
We observe that our system is never in the true infinite-size regime for 
$\s=1/4$, finite size effects are becoming evident for $\s=1/2$, even 
at early stages of evolution, and only for $\s\geq 3/4$ the boundary 
effects could be neglected.  
\begin{figure}
\centerline{\epsfxsize=8cm \epsfbox{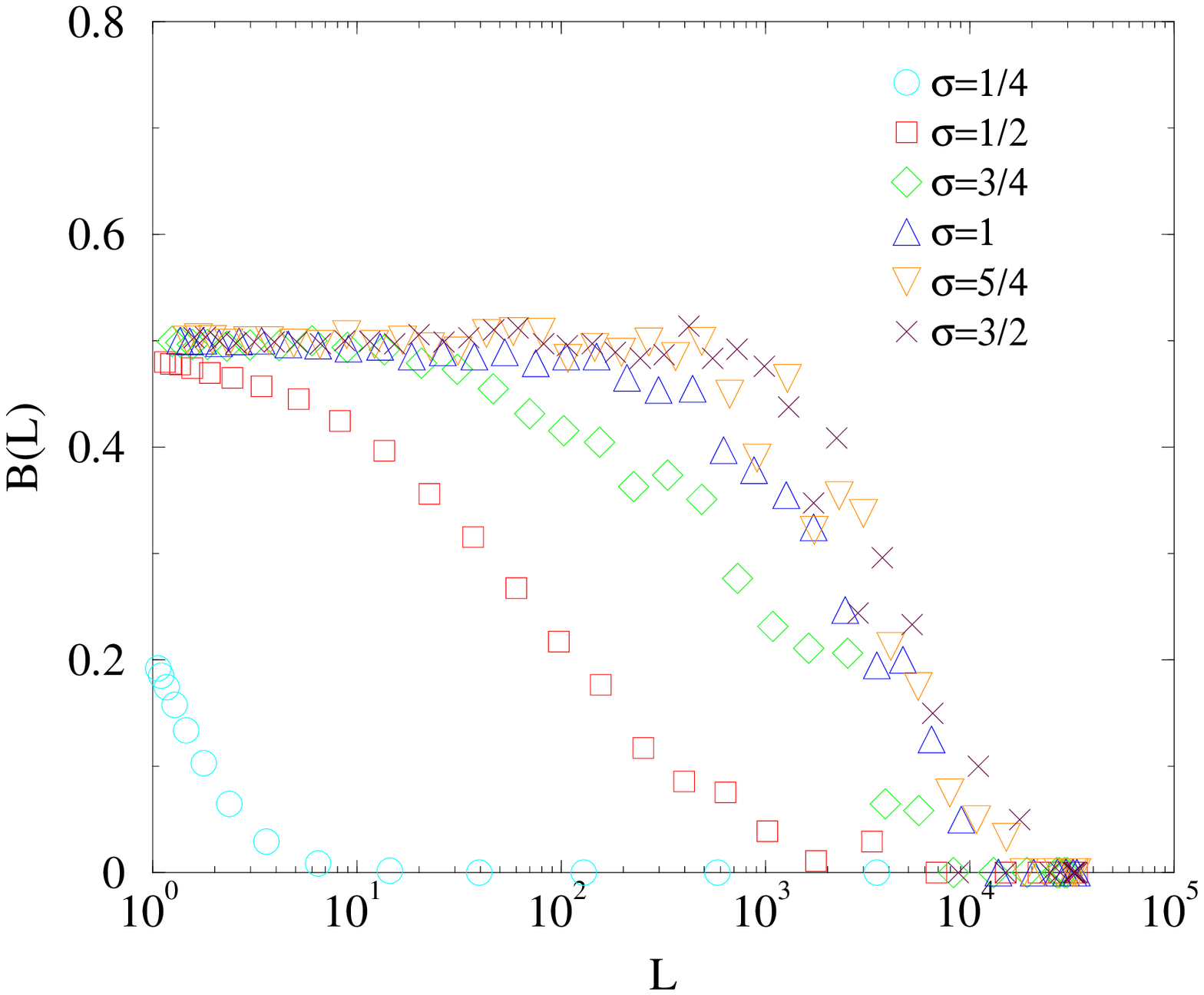}}
\noindent
{\small {\bf Fig.~4}. 
Number of domain walls $B(L)$ that move opposite to the direction 
prescribed by boundary effects vs. average domain length $L$
for various force exponents $\s$.} 
\end{figure}
Finally, we present a rough estimate of how big a system should be
for a particular 
value of $\s\ll 1$ to avoid significant finite-size effects.
We evaluate a force $F_{1-2}$,
exerted on a test domain wall by a dipole pair of 
neighboring domain walls,
\begin{equation}
F_{1-2}\approx {({1\over L})}^{\s}-
{({1\over2L})}^{\s}
= {({1\over L})}^{\s} (\s \ln{2}+{\cal O}({\s}^2))
\end{equation}
and compare it to the force $F_N$ exerted on 
the same test domain wall by a single domain wall at the the 
system boundary.
\begin{equation}
F_N \approx {({2\over N L})}^{\s}.
\end{equation}
Here $L$ and  $N$ are the typical domain length and the number of domains 
in the system. The boundary effects become significant
when these forces are of the same order. Hence for particular $\s$, the
minimum number of particles to avoid finite size effects $N_{min}$ is
\begin {equation}
N_{min}\approx 2{({1 \over \s\ln{2}})}^{1\over\s}
\end {equation}
We have observed (see Fig.~4) that for $\s=1/2$, $N_{min}\approx 10^5$.
Assuming the following parameterization of minimal size of the system 
for small $\s$, 
$N_{min}={({\mathrm constant} / \s)}^{1/ \s}$ and fitting it to
 $N_{min}(\s=1/2)= 10^5$, 
we obtain $N_{min}(\s=1/4)\approx1.6 \times 10^{11}$.
This is well beyond the limits of computational power available to us. 

In summary, we presented numerical evidence and a scaling argument 
suggesting the 
universality of persistent exponent for extremal model,
$\bar\theta=0.17507588\ldots$, for models with arbitrary force exponents
$\s>0$. We found that a deviation from scaling for persistence, 
that happens for small $\s$, is accompanied  by a similar deviation 
from scaling for a typical domain size $L(t)$ and is caused by finite size 
effects that cause crossover to a $\s=0$ coarsening scenario.
We estimated that in order to avoid boundary effects, the system 
size should grow as
${[{\cal O}(1/\s)]}^{1\s}$. 
A possible extension of this work is for higher dimensional systems, 
though the
duality between domain walls and spin dynamics that was extensively used 
for this work,
may not be so straightforward to apply.

The author would like to thank P. Krapivsky, A. Rutenberg, A. Hare, and
R. Hill for interesting discussions.

\end{multicols}

\end{document}